\documentclass[aps,pra,twocolumn,amsmath,amssymb,superscriptaddress,showpacs, 10pt]{revtex4-1}

\usepackage[utf8]{inputenc}
\usepackage{amsfonts}
\usepackage{amsmath}
\usepackage{amssymb}
\usepackage{amsthm}
\usepackage{fontenc}
\usepackage{graphicx}
\usepackage{xcolor}
\usepackage{textcomp}
\usepackage{epstopdf}
\usepackage{braket}
\usepackage{mathtools}
\usepackage{amsmath}
\usepackage{dcolumn}
\usepackage{multirow}

\begin{document}

\title{Photon-assisted transport in bilayer graphene flakes}
\author{D. Zambrano}
\author{L. Rosales}
\affiliation{Departamento de F\'isica, Universidad T\'ecnica Federico Santa Mar\'ia, Casilla 110-V, Valpara\'iso, Chile.}
\author{A. Latg\'e}
\affiliation{Instituto de F\'isica, Universidade Federal Fluminense, 24210-340 Niter\'oi-RJ, Brazil.}

\author{M. Pacheco}
\email{monica.pacheco@usm.cl}
\author{P. A. Orellana}
\affiliation{Departamento de F\'isica, Universidad T\'ecnica Federico Santa Mar\'ia, Casilla 110-V, Valpara\'iso, Chile.}
\date{\today}
\begin{abstract}

The electronic conductance of graphene-based bilayer flake systems reveal different quantum interference effects, such as Fabry-P\'erot resonances and sharp Fano antiresonances on account of competing electronic paths through the device. These properties may be exploited to obtain spin-polarized currents when the same nanostructure is deposited above a ferromagnetic insulator. Here we study how the spin-dependent conductance is affected when a time-dependent gate potential is applied to the bilayer flake. Following a Tien-Gordon formalism we explore how to modulate the transport properties of such systems via appropriate choices of the $ac$-field gate parameters. The presence of the oscillating field opens the possibility of tuning the original antiresonances for a large set of field parameters. We show that interference patterns can be partially or fully removed by the time-dependent gate voltage. The results are reflected in the corresponding weighted spin polarization which can reach maximum values for a given spin component. We found that differential conductance maps as functions of bias and gate potentials show interference patterns for different $ac$-field parameter configurations. The proposed bilayer graphene flake systems may be used as a frequency detector in the THz range.

\end{abstract}


\maketitle

%%%%%%%%%%%%%%%%%%%%%%%%%%%%%%%%%%%%%%%%%%%%%%%%%%%%%%%%%%%%%%
\section{Introduction}
%%%%%%%%%%%%%%%%%%%%%%%%%%%%%%%%%%%%%%%%%%%%%%%%%%%%%%%%%%%%%%

Graphene systems have been in the center of large amount of investigations and much effort is being done to overcome the challenges in providing concrete technological applications, exploring the interesting electronic properties of those two-dimensional realizations\cite{Geim2007,Fiori2014,Koppens2014}. The fast development on the graphene fundamental physics understanding has pushed the interest to explore new carbon-based hybrid nanostructures\cite{Nasibulin2007} and vertical van der Waals heterostructures\cite{Geim2013,Dean2012}. Efficient routes of fabrication and selective transfer of patterned graphene are required to provide optimum performance of the graphene devices\cite{Xudond2012,Wang2013}.

Recently\cite{Gehring2016}, graphene nanoconstrictions were fabricated using feedback controlled electro burning. Multiple transport features were provided such as multimode Fabry P\'erot (FP) interference patterns and Fano antiresonance states depending on the particular geometric details of the nanostructured systems. Experimental observation of FP interference in the conductance of a gate-defined cavity in a dual-gated bilayer graphene device has also been reported\cite{Varlet2014}. On the other hand, theoretical studies on the effects of $ac$-fields on the conductance and noise of carbon nanotubes and graphene nanoribbon devices in the FP regime \cite{Torres2009,Rocha2010} has been explored using the Tien-Gordon approach\cite{Tien1963,Platero2004,Foa2014}. Conductance interference patterns are shown to be modified by tuning of an $ac$-gate. For the case of nanoribbons the studies predicted a robust dependence of the resonator cavity responses on the nature of the graphene edges. Complex conductance profiles are also derived by joining an extra perturbation given by an applied magnetic field to the time-dependent field ($ac$-gate plate)\cite{Rocha2011}, providing an alternative way of coexistence of suppression state and regular oscillations, without the need of changing the frequency of the $ac$-field.

Previous studies on the electronic transport of bilayer graphene flakes have shown the system exhibits Fano antiresonances in the transmission \cite{Jhon2010, Jhon2011,Chico2012}. These properties has been exploited to obtain spin-polarized currents in a bilayer graphene flake deposited above a ferromagnetic insulator implying that the system may be utilized as a spin-filtering device \cite{orellana2013}. 

In this work, we investigate the effects of a time-dependent gate on the conductance and spin polarization in a bilayer graphene flake (BGF) in contact with a ferromagnetic insulator. We use a tight-binding Hamiltonian in the low energy approximation\cite{orellana2013} to obtain analytically the electronic transmission as a function of the length of graphene flake, the photon-assisted transport is investigated adopting the Tien-Gordon approximation, which properly describes photon-assisted inelastic tunneling events induced by a time-dependent potential. Our results show that the 
%%%%%%%%%%%%%%%%%%%%%%%%%%%%%%%%%%%%%%%%%%%%%%%%%%%%%%%%%%%%%%
% FIG 1
\begin{figure}[ht]
\centering
\includegraphics[width=1.00\linewidth]{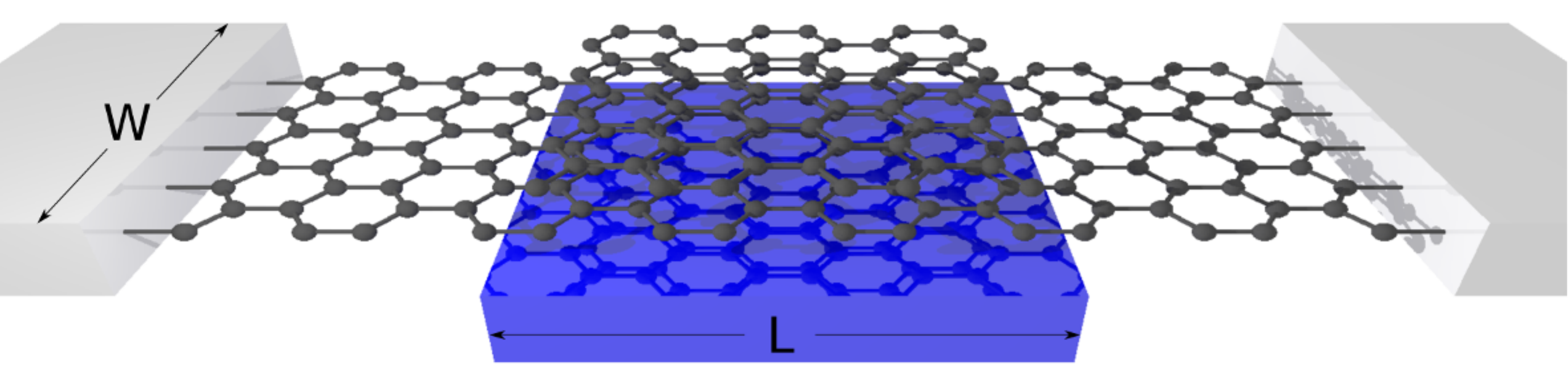}
\caption{(color online) Sketch of the model of a finite graphene flake of width $W$ deposited onto a graphene nanoribbon. The bilayer portion of the system (L) is placed on the proximity of a ferromagnetic material (blue). And an $ac$-field of intensity $V_{ac}$ and frequency $\omega$ is applied to the bilayer flake.}
\label{fig.montaje}
\end{figure}
presence of the time-dependent field provides a rich and interesting scenario for the electron transport in the BGF nanostructures. We found that calculated differential conductance maps depending on bias, $ac$-field and gate external potentials show complex Fabry-P\'erot like interference patterns for different $ac$-field parameter configurations. From the frequency values considered, we proposed that the BGF devices may be used as a frequency detector in the THz range.

%%%%%%%%%%%%%%%%%%%%%%%%%%%%%%%%%%%%%%%%%%%%%%%%%%%%%%%%%%%%%%
\section{The Model}
%%%%%%%%%%%%%%%%%%%%%%%%%%%%%%%%%%%%%%%%%%%%%%%%%%%%%%%%%%%%%%

In Fig. \ref{fig.montaje} we show schematically the studied graphene-based device composed of a flake of length $L = N/4$, $N$ being the number of carbon atoms along the flake. The flake is placed over an armchair nanoribbon (A-A stacked) connected to leads on both sides, given by perfect armchair ribbons of the same width $W$. Although it is known that A-B stacking is more stable for graphene layered systems, the A-A stacking has also been reported experimentally in few-layer systems\cite{liu2009}. Since both BGF stacking-like exhibit antiresonance phenomena due to the coupling of discrete and extended states,  A-A stacking was chosen over the A-B configuration to simplify the analytic formulation of the problem, without loss of generality. The scattering region is placed over a ferromagnetic material that induces an exchange splitting due to the proximity effects. In what follows we use the value of $\Delta_\text{ex}=0.02 \gamma_0=0.054 eV$ for the exchange splitting\cite{Lee2010,haugen2008,Wang2015}.

The Hamiltonian of the system is written in the single $\pi$-band tight binding formalism as a sum of the leads contributions (left and right), and the central part 
\begin{equation}
	H = H_L + H_R + H_C\,\,,
	\label{hami}
\end{equation}
where
\begin{equation}
	H_{L,R} = -\gamma_0\sum_{<i,j>,\sigma}c_{i,\sigma}^\dagger c_{j,\sigma}\,\,,
\end{equation}
with $c_{i,\sigma}^\dagger$ and $c_{i,\sigma}$ being the creation and annihilation operators of an electron with spin $\sigma$ in site $i$, where $<i,j>$ means that the sum is over first neighbors. For graphene systems the energy hopping between first neighbors is usually taken as $\gamma_0\approx 2.7~eV$ \cite{DFT}. The Hamiltonian of the central part includes a time-dependent gate potential $V(t)$, and the exchange potential energy,
\begin{eqnarray}
H_C = &-&\gamma_0\sum_{<i,j>,\sigma,\alpha}c_{i,\sigma}^{\alpha\dagger} c_{j,\sigma} ^{\alpha} -\gamma_1\sum_{i,\sigma}\left( c_{i,\sigma}^{u\dagger} c_{i,\sigma}^{d} + \text{h.c.} \right)\nonumber\\
      &+&\sum_{i,\sigma,\alpha}({ \varepsilon_i +eV(t)}+\sigma\Delta_{\text{ex}})c_{i,\sigma,}^{\alpha\dagger} c_{i,\sigma}^{\alpha} \,\,,
\end{eqnarray}
where $\varepsilon_i$ is the onsite energy and $\gamma_1=0.1\gamma_0$ being the hopping energy between first neighbors atoms from the upper and lower layers in the central region of the device. The index $\alpha$ takes values $u$ and $d$ for the upper and lower ribbons in the scattering region, $\sigma$ is the spin index and takes values $\pm 1$. The time-dependent gate potential is given by $V(t) =V_{g}+ V_{\text{ac}}\cos(\omega t)$, with $V_{g}$ being a constant gate potential, $V_{ac}$ the $ac$-field intensity, and $\omega$ the field frequency.

To calculate the transport properties of the system in the presence of the considered homogeneous time-dependent gate we adopt a Tien-Gordon-like theory\cite{Tien1963} in which it is assumed that both spatial and time dependencies of the Floquet states can be factorized. Within this approximation the time average spin-polarized current can be written as 
\begin{widetext}
\begin{equation}
	<J_{\sigma}> = \frac{e}{h}\sum_{m=-\infty}^\infty {\mathcal J}^{2}_m\left( \frac{eV_{\text{ac}}}{\hbar\omega}\right) \int dE (f_L - f_R) T_{\sigma}(E+m\hbar\omega,V_{g})\,\,,
	\label{current}
\end{equation}
\end{widetext}
where $f_{L(R)}$ is the Fermi distribution of the left (right) lead and ${\mathcal J}_m$ are the Bessel functions of first kind. The integral is evaluated between the left and right chemical potentials $\mu_{L,R}= \varepsilon_{F} \pm \frac{eV_b}{2}$, $\varepsilon_{_F}$ being the Fermi energy and $V_{b}$ the bias potential. Here $T_{\sigma}(E)$ is the spin-dependent transmission function in the absence of the $ac$-field which depends on the gate voltage $eV_g$ and the exchange potential $\sigma\Delta_\text{ex}$.

The spin-dependent transmission can be analytically obtained in the single-mode approximation as described in Ref.\cite{orellana2013} in which the bilayer flake is reduced to an equivalent problem of two coupled linear chains, and is given by (see appendix \ref{A1})
\begin{widetext}
\begin{equation}
	T_{\sigma}(E) = \frac{\left( 4 - \frac{E^2}{\gamma_0^2} \right)Q_{\sigma}^2 }{\left[ \left( Q_{\sigma}^2 - R_{\sigma}^2 - 1 \right)\frac{E}{2\gamma_0} + 2R_{\sigma} \right]^2 + \left( Q_{\sigma}^2 - R_{\sigma}^2 + 1 \right)^2 \left( 1 - \frac{E^2}{4\gamma_0^2} \right) }\text{,}
	\label{T_previous}
\end{equation}
\end{widetext}
with $Q_{\sigma}$ and $R_{\sigma}$ are given by,
\begin{eqnarray*}
	Q_{\sigma} &=& -\frac{1}{2}\left[ \frac{1}{U_N(x^{+}_{\sigma})} + \frac{1}{U_N(x^{-}_{\sigma})} \right]\text{,} \\
\\
	R_{\sigma} &=& \frac{1}{2} \left[\frac{U_{N-1}(x^{+}_{\sigma})} {U_N(x^{+}_{\sigma})} + \frac{U_{N-1}(x^{-}_{\sigma})}{U_N(x^{-}_{\sigma})} \right] \text{,}
\end{eqnarray*}
$U_N(x^{\pm}_{\sigma})$ being the second-kind Chebyshev polynomials with $x^{\pm}_{\sigma}=arccos (\left(E - eV_{g} - \sigma\Delta_{\text{ex}}\mp\gamma_1\right)/2\gamma_0)$. 

The spin-dependent conductance is written as follows
\begin{equation}
	G^{ac}_{\sigma} = G_0\sum_{m=-\infty}^\infty {\mathcal J}^{2}_m\left( \frac{eV_{\text{ac}}}{\hbar\omega}\right) T_{\sigma,m}\left(\varepsilon_F+m\hbar\omega \right)\text{,}
	\label{conductance}
\end{equation}
where $G_0 = e^2/h$ is the quantum conductance per spin. 

The spin polarization of the electric current is defined as a weighted quantity \cite{SOB}, given by
\begin{equation}
P_{\sigma} = \frac{\left|G_{\uparrow} - G_{\downarrow}\right|}{\left|G_{\uparrow} + G_{\downarrow}\right|}G_{\sigma}\,\,.
\label{polarization}
\end{equation}

%%%%%%%%%%%%%%%%%%%%%%%%%%%%%%%%%%%%%%%%%%%%%%%%%%%%%%%%%%%%%%
\section{Results}
%%%%%%%%%%%%%%%%%%%%%%%%%%%%%%%%%%%%%%%%%%%%%%%%%%%%%%%%%%%%%%
In what follows we show spin-dependent conductance $G_\sigma$ and weighted spin polarization $P_\sigma$ results for ferromagnetic BGFs with different scattering regions (bilayer flake). The systems
%%%%%%%%%%%%%%%%%%%%%%%%%%%%%%%%%%%%%%%%%%%%%%%%%%%%%%%%%%%%%%
% FIG 2
\begin{figure}[h]
\centering
\includegraphics[width=1.00\linewidth]{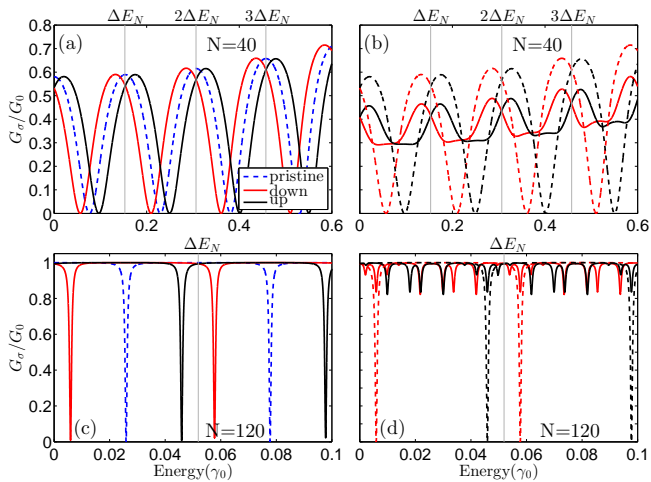} 
\caption{(color online) Spin-dependent conductance versus energy for two BGFs with lengths $N=40$ in (a) and (b) panels and $N=120$ in (c) and (d). Full red/black curves denote spins down/up. The systems are probed by an $ac$-field of intensity $eV_{ac}=0.046 \gamma_0$ and frequencies $\hbar\omega=\Delta E_{40}= 0.153\gamma_0$ in (a), $\hbar\omega=\Delta E_{120}=0.052\gamma_0$ in (c);     dashed blue lines are used for the pristine system ($V_{ac}=0$ and $\Delta_\text{ex}=0$). In (b) and (d) the systems are probed with the same frequency $\hbar\omega=0.012\gamma_0$ and dashed red/black lines stand for $V_{ac}=0$.}
\label{fig2}
\end{figure}
are photon-assisted by a time-dependent $ac$-field with energies going from $\hbar\omega = 0.001\gamma_0$ to $0.4\gamma_0$, corresponding approximately to frequencies between $650$ GHz-$260$ THz. 

Here we consider two particular device lengths: $N = 40$ and $120$, $N$ being the number of carbon atoms along the flake ($N = 4L$). The lengths were actually chosen as typical examples of different transport regimes when the $ac$-potential is null \cite{orellana2013}: for $N=40$ the system shows the characteristic FP oscillations in the conductance while the $N=120$ device exhibits sharp Fano antiresonances in a plateau with maximum transmission, providing higher spin-polarized conductances.

In the following, we define the energy period of the conductance oscillations of the unperturbed system as $\Delta E_{_N}$. The dependence of $\Delta E_{_N}$ with $N$ can be derived from the transmission probability given in Eq. (\ref{T_previous}) in the absence of the $ac$-field and the ferromagnetic insulator. Starting directly from second-kind Chebyshev polynomials we found an analytic expression for $\Delta E_{_N}$,
\begin{equation}
\Delta E_{_N} = 2\sin\left( \frac{\pi}{N+1} \right) \cos\left( \frac{\pi}{2N+2} \right)\gamma_0\,\,.
\label{fig_DE_fit}
\end{equation}

In a first-order approximation, for $N>>1$, $\Delta E_{_N} \approx \frac{2\pi}{N+1}\gamma_0$, in accordance with the continuum model\cite{Jhon2010}. The maxima of the conductance results related to the unperturbed systems (dashed blue curves) are marked in Fig. \ref{fig2} with integer multiples of $\Delta E_{_N}$, following Eq. (\ref{fig_DE_fit}). 

%%%%%%%%%%%%%%%%%%%%%%%%%%%%%%%%%%%%%%%%%%%%%%%%%%%%%%%%%%%%%%
% FIG 3
\begin{figure}[h]
\centering
\includegraphics[width=1.00\linewidth]{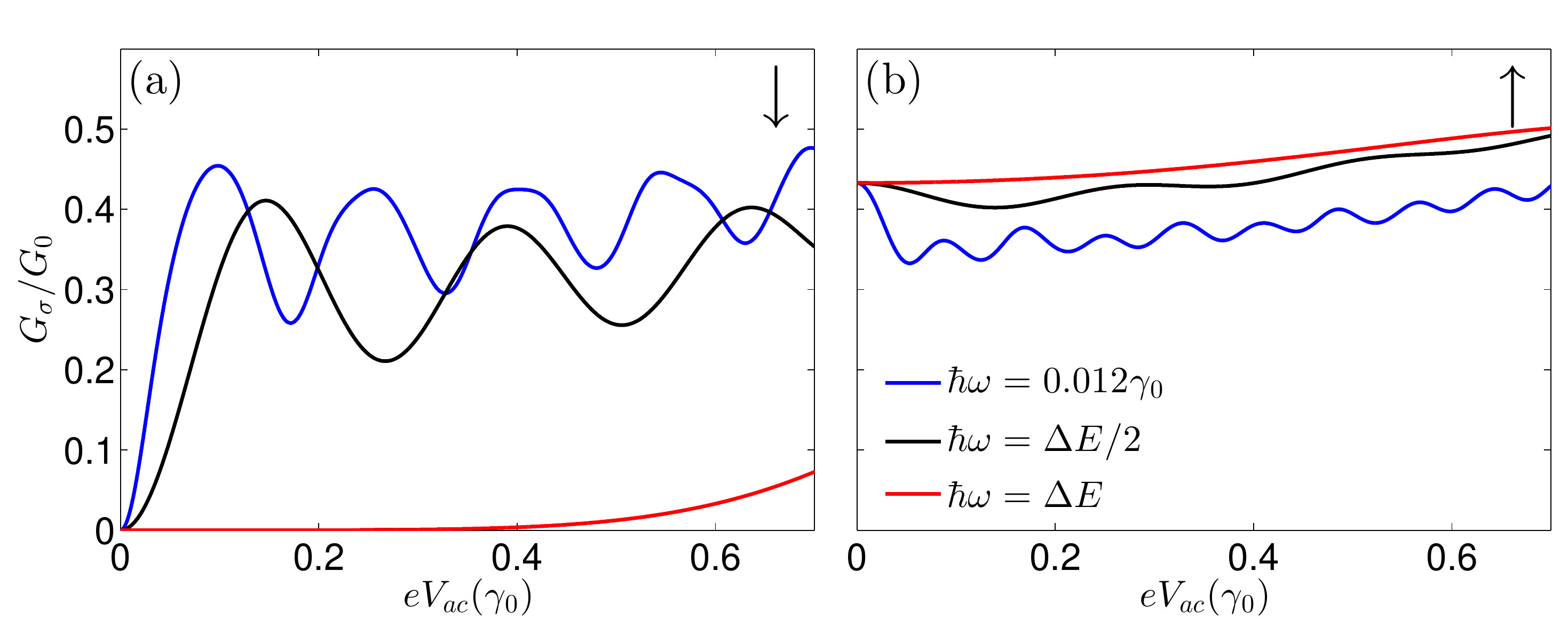} 
\caption{(color online) Spin-dependent conductance as a function of the intensity of the $ac$-field intensity ($eV_\text{ac}$) for a BGF of length $N=40$, for different values of the $ac$-field frequency ($\hbar\omega$). The curves are calculated for a fixed energy $E = \Delta E_{_N}/2 - \Delta_\text{ex} \approx 0.06\gamma_0$ corresponding to a minimum of the spin down conductance.}
\label{fig3}
\end{figure}

The effects of the $ac$-field on the spin-dependent conductance are shown in Fig. \ref{fig2} for the two systems considered; $N=40$ (top panels) and $N=120$ (bottom panels). Spin down and up are shown in red and black curves. For comparison, the conductance results of the unperturbed system ($V_{ac}=0$ and $\Delta_\text{ex}=0$) are also shown in dashed blue curves in Figs. \ref{fig2}(a) and (c). The $ac$-field intensity is fixed at $eV_{ac}=0.046 \gamma_o$ and we investigate the effect of tuning the $ac$-field frequencies. For the cases in which $\hbar\omega$ coincides with the energy period of the conductance oscillations of the unperturbed system, $\hbar \omega=\Delta E_{40}= 0.153 \gamma_0$ for $N=40$ [Fig. \ref{fig2} (a)] and $\hbar \omega=\Delta E_{120}=0.052\gamma_0$ for $N=120$ [Fig. \ref{fig2} (c)] the conductance is completely unresponsive to the presence of the $ac$-field. This result can be interpreted as due to the  stroboscopic or wagon-wheel condition predicted in these kind of quantum devices \cite{Rocha2010,Rocha2011,Gehring2016}. The physical scenario is different for frequencies other than integer multiples of $\Delta E_{_N}$, as shown in Fig. \ref{fig2} (b) and (d) for $\hbar\omega = 0.012\gamma_0$. In this $ac$-field regime, the amplitude of the conductance resonances decreases and new oscillatory features appears as the Fermi energy is changed. Also, the Fano antiresonances are completely removed in the spin-dependent conductances compared with the case of null $ac$-field (dashed curves). For both systems the average value of the conductance is the corresponding to the static case (null $ac$-field). 

%%%%%%%%%%%%%%%%%%%%%%%%%%%%%%%%%%%%%%
%FIG4
\begin{figure}[h]
\centering
\includegraphics[width=1.\linewidth]{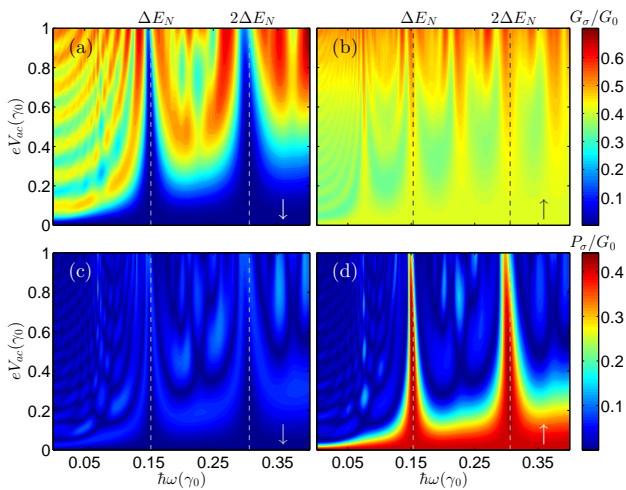}
\caption{(color online) Contour plots of the conductance (upper panels) and weighted spin-polarization (bottom panels) for an $N=40$ BGF as function of the $ac$-field parameters, $eV_{ac}$ and $\hbar\omega$. The energy is fixed at $E = \Delta E_{40}/2 - \Delta_\text{ex} \approx 0.06\gamma_0$, (a) and (c) correspond to spin down, and (b) and (d) to spin up.}
\label{fig.GandP40_3}
\end{figure}

The spin-dependent conductance as a function of the $ac$-field intensity exhibits different features depending on the $ac$-frequency value, as shown in Fig. \ref{fig3} for a BGF of length $N=40$. The curves are calculated for a fixed energy $E = \Delta E_{40}/2 - \Delta_\text{ex} \approx 0.06\gamma_0$, corresponding to a minimum of the spin down conductance, and for different values of the field frequency $\hbar\omega=0.012\gamma_0$ (blue curves), $\Delta E_{40}$ and $\Delta E_{40}/2$ (black and red curves, respectively). The predicted stroboscopic effects are evidenced when $\hbar\omega=\Delta E_{40}$, for a large range of $ac$-field intensity ($eV_\text{ac}$) values and for both spin components. The same condition is also realized for $\hbar\omega=\Delta E_{40}/2$ in the case of spin up conductance (Fig. \ref{fig3}(b)) while the spin down component exhibits an oscillatory response. For the low frequency considered, $\hbar\omega=0.012\gamma_0$ (blue curves), FP oscillations as a function of the $eV_{ac}$ intensity are clearly observed. It is relevant to note that down and up spin components of the conductance show distinct FP periods, i.e., $\Delta E_{40}$ and $\Delta E_{40}/2$, respectively.

The effects of the intensity and frequency of the $ac$-field on the spin-dependent conductance and may be better analyzed by means of the contour plots displayed in Fig. \ref{fig.GandP40_3} for $E = \Delta E_{40}/2 - \Delta_{\text{ex}} \approx 0.06\gamma_0$, as in the previous figure. The down and up spin conductance results are shown in Fig. \ref{fig.GandP40_3} (a) and (b), respectively. As anticipated, the results for the spin down conductance exhibit minimum values for $\hbar\omega$ multiples of $\Delta E_{_N}$, independently of the intensity of the $ac$-field. In the low frequency regime, $\hbar\omega<<\Delta E_{_N}$, the FP oscillations as a function of the $eV_{ac}$ intensity are highlighted. For the spin up direction, as the fixed energy does not correspond to a minimum in the conductance due to the magnetic energy shift, zero conductance regions (dark blue in the top-left map) are not found, although the interference patterns in the conductance as a function of the $ac$-field parameters are still clear. However, in this case a stroboscopic effect can also be discerned at $\hbar \omega $ semi-multiples of $\Delta E_{_N}$ (Fig. \ref{fig.GandP40_3} (b)). This particular feature is similar to that found in Ref. \onlinecite{Rocha2011} where the effects of the interplay between static magnetic fields and an $ac$-field on carbon-based resonant cavities were investigated. A simple explanation can be given considering the photonics excitations classified into two different families according to their parity. 

%%%%%%%%%%%%%%%%%%%%%%%%%%%%%%%%%%%%%%%%%%%%%%%%%%%%%%%%%%%%%%
% FIG 5
\begin{figure}[ht]
\centering
\includegraphics[width=1.00\linewidth]{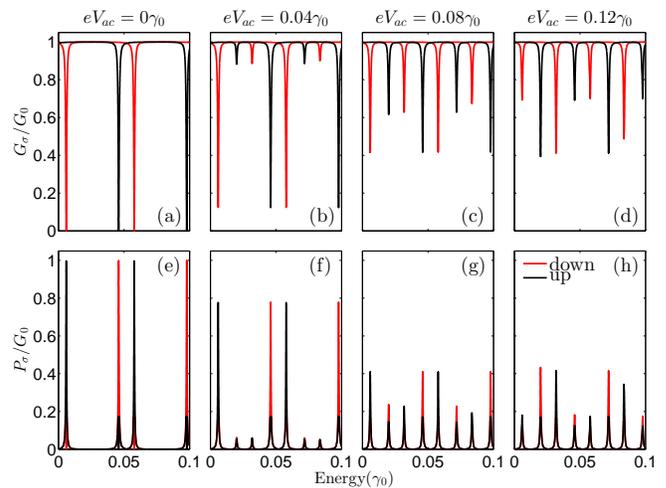}
\caption{(color online) Spin-dependent conductance (a-d) and weighted spin polarization (e-h) versus energy for a 120-GNR. The $ac$ frequency is fixed at $\hbar\omega = 3\Delta E_{120}/2 \approx 0.08(\gamma_0)$ and the $ac$-field intensity is increasing: (a) $eV_{\text{ac}} = 0$, (b) $0.04\gamma_0$, (c) $0.08\gamma_0$, and (d) (e) $0012\gamma_0$. Red and black curves stand for spin down and up respectively.}
\label{fig.PT_compare_120}
\end{figure}

%%%%%%%%%%%%%%%%%%%%%%%%%%%%%%%%%%%%%%%%%%%%%%%%%%%%%%%%%%%%%%
% FIG 6
\begin{figure}[h]
\centering
\includegraphics[width=1.05\linewidth]{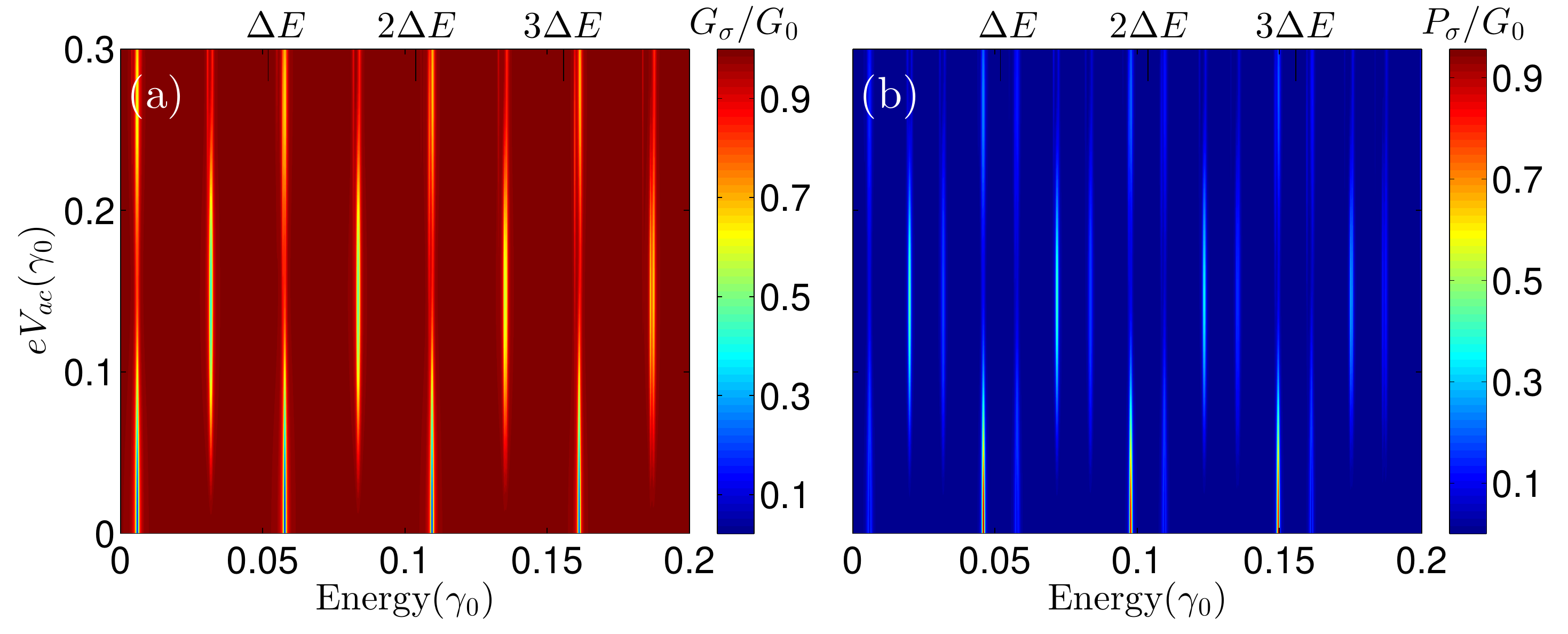}
\caption{(color online) Contour plots for the spin down conductance (a), and for weighted spin polarization (b), as a function of the $ac$-field intensity. The parameters are the same as in figure \ref{fig.PT_compare_120}.}
\label{fig.G_vs_E_V_120}
\end{figure}

The weighted spin-polarization is also investigated in terms of the $ac$-field parameters, as shown in Fig. \ref{fig.GandP40_3} (c) and (d). The results show that the oscillating field may actually tune the spin polarization, from low to high values. For this particular parameter configuration the up weighted spin-polarization achieves a maximum of 45\%. 

In the following we analyze the spin-dependent conductance and spin polarization for the $N=120$ BGF under different $ac$-field parameter regimes. The panels shown in Fig. \ref{fig.PT_compare_120} correspond to results for different values of the $ac$-field intensity ($0-0.12\gamma_0$) and a fixed $ac$-field frequency ($\hbar\omega=3\Delta E_{120}/2 \approx 0.08\gamma_0$). It can be noted that the presence of the oscillating field gives rise to new quasi-antiresonances in the conductance at energies equal to semi multiples of $\Delta E_{_N}$, with the corresponding shift $\pm\Delta_\text{ex}$ for up and down components. At the same time, the amplitudes of the antiresonances, pinned at an integer multiple of $\Delta E_{_N}$, are reduced as shown in Fig. \ref{fig.PT_compare_120} (a)-(d). The weighted spin polarization exhibits the same behavior as the conductance, as shown in Figs. \ref{fig.PT_compare_120} (e)-(f). For the particular value of $eV_{\text{ac}}=0.093\gamma_0$ both peaks related to up and down spins are equal, achieving $50\%$ of spin polarization. 

The contour plots in Fig. \ref{fig.G_vs_E_V_120} (a) and (b) show the spin-dependent conductance and weighted spin-polarization for spin down, respectively, as a function of the $ac$-field intensity and the Fermi energy. In the static case ($V_{ac}=0$), the antiresonances in the conductance are located at energies equal to integer multiples of $\Delta E_{_N}\pm \Delta_\text{ex}$. As the $eV_{ac}$ field intensity increases, an oscillatory regime is established and quasi anti-resonances (harmonic) appear at energies equal to half integer multiples of $\Delta E_{_N}/2 \pm \Delta_\text{ex}$. These anti-resonances oscillate out of phase with respect to the original ones as shown in Fig. \ref{fig.G_vs_E_V_120} (a). It is also interesting to investigate the opening and closing of the spin-polarization channels as the $eV_{ac}$ field intensity is continuously changed for the characteristic energies discussed. For the energies corresponding to Fano antiresonances, at integer multiples $\Delta E_{120}$, optimum values up to 95\% of the weighted spin polarization are obtained (Fig. \ref{fig.G_vs_E_V_120} (b)), in a large range of the $ac$-field intensities. 

%%%%%%%%%%%%%%%%%%%%%%%%%%%%%%%%%%%%%%%%%%%%%%%%%%%%%%%%%%%%%% exhibits the same  as the conductance
% FIG 7
\begin{figure}[ht]
\centering
\includegraphics[width=1.00\linewidth]{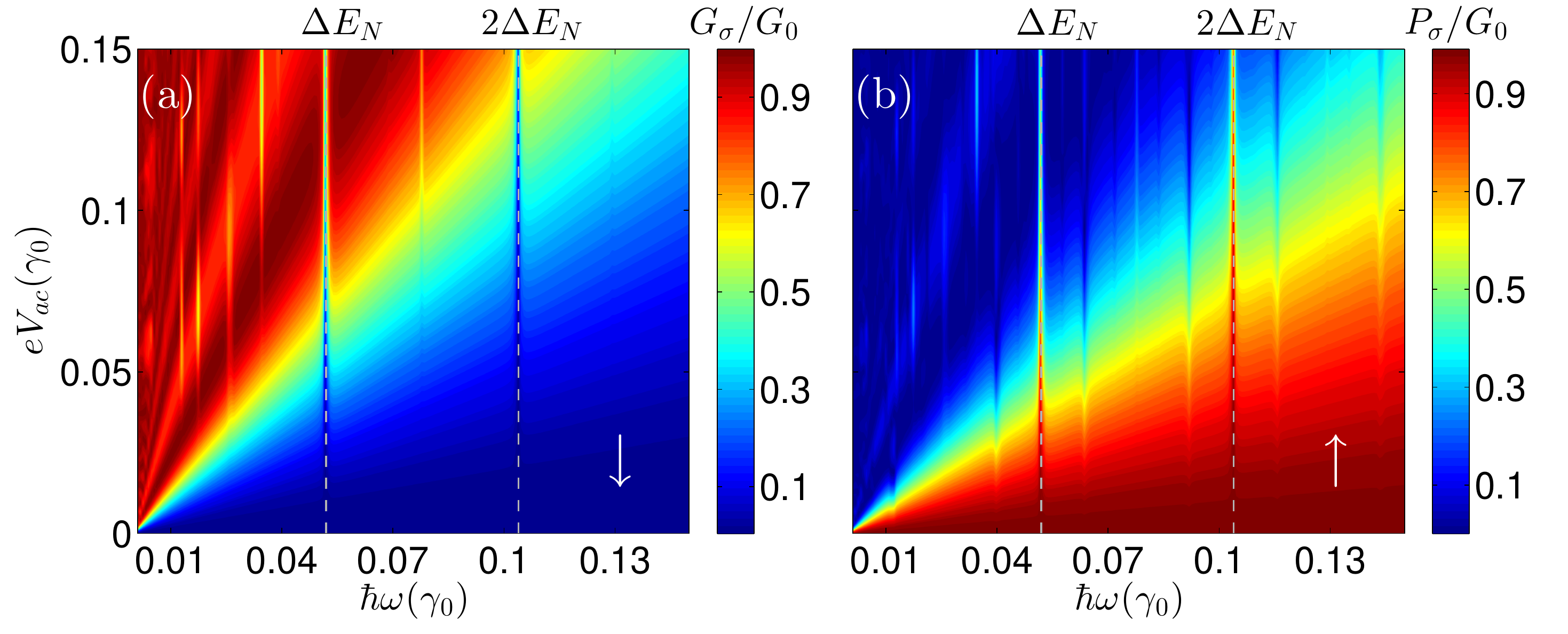}
\caption{(color online) Contour plots of the (a) spin down conductance and (b) weighted spin up polarization as a function of the $ac$-field frequency $\hbar\omega$ and the $ac$-field potential $eV_{ac}$, for a $N=120$ BGF. The energy is fixed at $E = \Delta E_{120}/2 - \Delta_{\text{ex}} \approx 0.006\gamma_0$.}
\label{fig.GandP120_2}
\end{figure}

The effects of the strength and frequency of an $ac$-field on the spin-dependent conductance and the weighted spin polarization for a $N=120$ BGF are displayed in the contour plots in Fig. \ref{fig.GandP120_2}. The results are calculated for an energy $E=\Delta E_{120}/2 - \Delta_{\text{ex}} \approx 0.006\gamma_0$, corresponding to the first minimum of the spin-down conductance in Fig. \ref{fig2}(c) (red line). The spin down conductance results are shown in Fig. \ref{fig.GandP120_2} (a). Contrary to the $ac$-field response of the $N=40$ bilayer flake system, in the low frequency regime ($\hbar\omega<<\Delta E_{_N}$), the FP oscillations as a function of the $eV_{ac}$ magnitude are suppressed for this $N=120$ device. The Fano antiresonances, however, are still present for the corresponding stroboscopic frequencies $\hbar\omega$ integer multiples of $\Delta E_{_N}$. The presence of the $ac$-field opens the possibility of tuning the primary antiresonances for a large set of field parameters. For spin down conductance this is evident from the triangular blue region in the contour plot (a). These results are reflected in the corresponding weighted spin polarization in which the up component (panel (b)) exhibits maximum polarization values (99\%) in the same range of $eV_{ac}$ intensities and $\hbar\omega$ frequencies. Same results are obtained if we analyze the spin up conductance and weighted spin-down polarization for an energy corresponding to the up-component Fano antiresonance.
 
%%%%%%%%%%%%%%%%%%%%%%%%%%%%%%%%%%%%%%%%%%%%%%%%%%%%%%%%%%%%%%
% FIG 8
\begin{figure}[ht]
\centering
\includegraphics[width=1.0\linewidth]{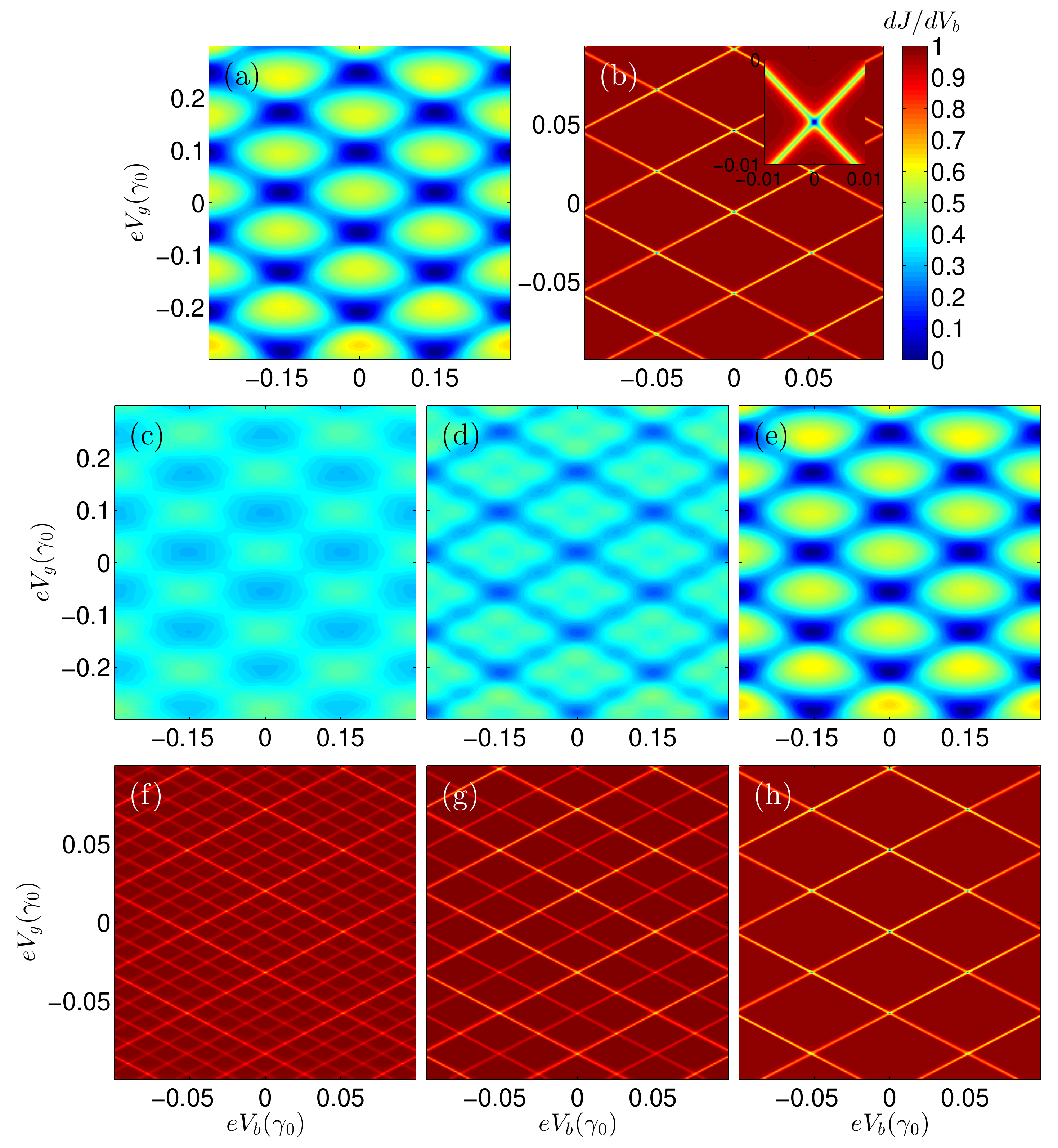}
\caption{(color online) Differential conductance contour plots as a function of bias and gate voltages. Top panels: null $ac$-field for $N=40$ (left) and $N=120$ (right). Middle panels: $eV_\text{ac}=0.172\gamma_0$ for $N=40$; Bottom panels: $eV_\text{ac}=0.046 \gamma_0$ for $N=120$ and from left to right frequencies: $\hbar\omega = \Delta E_{_N}/4$, $\Delta E_{_N}/2$ and $ \Delta E_{_N}$, for spin down.}
\label{fig.8}
\end{figure}

Next, we investigate the BGF devices under applied bias $V_b$ and gate $V_{g}$ voltages. In the top panels of Fig. \ref{fig.8} it is shown the results of the differential conductance for null $ac$-field for $N=40$ (left) and $N=120$ (right). The expected FP quantum interference patterns appear in the characteristic transport diagram for the bias-dependent conductance. For the $N=120$ BFG device the presence of the Fano antiresonances are clearly revealed in the conductance by very well delineated diamonds. These are highlighted in the inset of Fig. \ref{fig.8} (b) by a bright blue point corresponding to zero conductance value. The effects of the $ac$-field on the FP patterns are shown in the middle and bottom panels for the $N=40$ ($eV_\text{ac}=0.172\gamma_0$) and $N=120$ ($eV_\text{ac}=0.046\gamma_0)$ BGF devices, respectively. The results for the $ac$-field frequencies $\hbar\omega=\Delta E_{_N}/4$, $\Delta E_{_N}/2$, and $\Delta E_{_N}$, are depicted from left to right. Similar results are found for the other spin direction, with the corresponding energy shift in the gate voltage.  

These $ac$ parameters have been chosen to highlight the behavior of the different regimes: (1) the low frequency regime [(c) and (f)] for which the amplitude of the FP oscillations is strongly reduced even though the interference patterns are still outlined; (2) frequencies in the stroboscopic condition[(e) and (h)] for which the differential conductance recovers the original FP resonances corresponding to the static regime, and (3) frequencies equal to semi multiples of the stroboscopic frequency [(d) and (g)] that show interesting interference patterns exhibiting new periodic structures in the FP diamonds of the conductance maps. The studied systems cover different new features of photon-assisted quantum transport occurring into the GHz and THz frequency spectra. Actually, measurements of the quantum shot noise in a graphene nanoribbon subjected to electromagnetic THz radiation were recently reported\cite{Parmentier2016}, indicating that the studied bilayer graphene flake system may be realized and proposed as THz frequency detector devices.

%%%%%%%%%%%%%%%%%%%%%%%%%%%%%%%%%%%%%%%%%%%%%%%%%%%%%%%%%%%%%%
\section{Summary}
\begin{figure}[ht]
\centering
\includegraphics[width=1.0\linewidth]{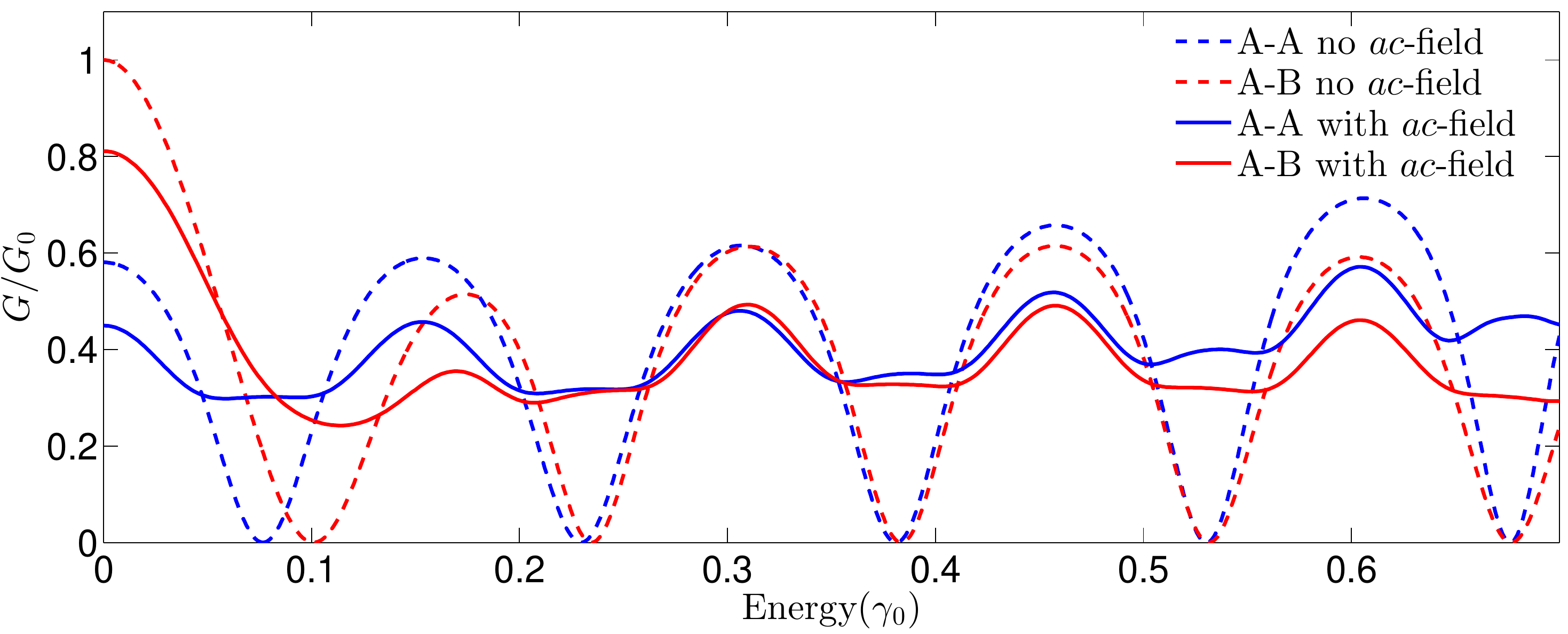}
\caption{(color online) BGF conductance versus energy for A-A (blue curves) and A-B (red curves) stacking. Conductances for the pristine systems are marked with dashed lines and in the presence of an $ac$-field with full lines. The parameters are the same as in Fig. \ref{fig2}(b), with $\Delta_\text{ex} = 0$.}
\label{fig.9}
\end{figure}

%%%%%%%%%%%%%%%%%%%%%%%%%%%%%%%%%%%%%%%%%%%%%%%%%%%%%%%%%%%%%%
We have studied the spin-dependent electron transport in graphene-based bilayer flake systems in contact with a ferromagnetic insulator when an $ac$-field gate potential is applied. We have obtained an analytic expression for the conductance using and A-A stacking for the flake. Nevertheless, the antiresonance phenomena reported here can be found in A-B stacking as well (see Fig. \ref{fig.9}). In the static regime 
%%%%%%%%%%%%%%%%%%%%%%%%%%%%%%%%%%%%%%%%%%%%%%%%%%%%%%%%%%%%%%
% FIG 9
these systems have revealed different quantum interference effects in the conductance as a function of the flake length\cite{orellana2013} such as: Fabry-P\' erot resonances and sharp Fano antiresonances which appear because of the competing electronic paths taking place in the systems. The time-dependent transport properties have been calculated in the Tien-Gordon formalism where the time-averaged spin-polarized current is obtained in terms of the electronic transmission for null $ac$-field. We have shown that interference patterns in the conductance can be tuned, being partially or fully removed by the $ac$-field gate, with a proper selection of parameters. The Fano antiresonance conductance spectra can be strongly modified for particular values of the field frequency for which the period of the antiresonances is half of the original and its amplitudes oscillate as a function of the $ac$-field intensity. In this way the presence of the $ac$-field opens the possibility of tuning the original antiresonances in a large set of field parameters. These results are reflected in the corresponding weighted spin polarization which can reach maximum values for a given spin component. We found that calculated differential conductance maps depending on bias and gate external potentials show complex Fabry-P\' erot like interference patterns for different $ac$-field parameter configurations. From the frequency values considered, we conclude that the proposed BGF devices may be used as a frequency detector in the THz range.

\acknowledgments{The authors acknowledge financial support from FONDECYT under Grants 1140571, 1151316 $\&$ 1140388 and from CONICYT under Grant PAI-79140064 and Act-1204. AL thanks the financial support of FAPERJ under Grant E-26/102.272/2013, CNPq and INCT de Materiais de Carbono.}

\appendix
    \section{Transmission function}\label{A1}

From the Hamiltonian, eq. (\ref{hami}), and within the tight-binding approximation, the probability amplitude of finding an electron at an atom $A(B)$ in a site $j(k)$ in a double layer graphene nanoribbon is given by,
\begin{eqnarray}
	i\hbar \frac{\partial}{\partial t} \psi_{j,k,\sigma}^{A,\alpha} &=& \gamma_0\left( \psi_{j,k,\sigma}^{B,\alpha} + \psi_{j-1,k-1,\sigma}^{B,\alpha} + \psi_{j-1,k+1,\sigma}^{B,\alpha} \right) \nonumber\\
	&+& \gamma_1\psi_{j,k,\sigma}^{A,\overline{\alpha}} + \left( \sigma\Delta_{\text{ex}} + eV(t) \right) \psi_{j,k,\sigma}^{A,\alpha} \text{,} \nonumber\\
	i\hbar \frac{\partial}{\partial t} \psi_{j,k,\sigma}^{B,\alpha} &=& \gamma_0\left( \psi_{j,k,\sigma}^{A,\alpha} + \psi_{j+1,k-1,\sigma}^{A,\alpha} + \psi_{j+1,k+1,\sigma}^{A,\alpha} \right) \nonumber\\
	&+& \gamma_1\psi_{j,k,\sigma^*}^{B,\overline{\alpha}} + \left( \sigma\Delta_{\text{ex}} + eV(t) \right) \psi_{j,k,\sigma}^{B,\alpha}  \text{,} \label{eq.motion}
\end{eqnarray}
where all the wave functions $\psi$ are time dependent, $\psi(t)$, and $V(t) = V_g + V_\text{ac}\cos(\omega t)$.

For the direction perpendicular to the current, the solutions can be written as a combination of plane waves of the form $e^{\pm imq}$, and then equation (\ref{eq.motion}) is reduced to the equivalent problem of two coupled linear chain. Then the reflection and transmission coefficients are obtained, leading to eq. (\ref{T_previous}).


\begin{thebibliography}{0}

\bibitem{Geim2007} A. K. Geim and K. S. Novoselov, Nat. Mater. \textbf{6}, 183 (2007).

\bibitem{Fiori2014} G. Fiori, F. Bonaccorso, G. Iannaccone, T. Palacios, D. Neumaier, A. Seabaugh, S. K. Banerjee and L. Colombo, Nature Nanotech. 9, 768 (2014).

\bibitem{Koppens2014} F. H. L. Koppens, T. Mueller, Ph. Avouris, A. C. Ferrari, M. S. Vitiello, and M. Polini, Nature Nanotech. 9, 780 (2014).

\bibitem{Nasibulin2007} A. G. Nasibulin et al, Nature Nanotechnology \textbf{2}, 156 (2007).

\bibitem{Geim2013} A. K. Geim and I. V. Grigorieva, Nature \textbf{499}, 419 (2013).

\bibitem{Dean2012} C. Dean, A. F. Young, L. Wang, I. Meric, G.-H. Lee, K. Watanabe, T. Taniguchi, K. Shepard, P. Kim, and J. Hone, Solid State Commun. \textbf{152}, 1275 (2012).

\bibitem{Xudond2012} C. Xu-Dong, L. Zhi-Bo, J. Wen-Shuai, Y. Xiao-Qing, F. Xing, W. Peng, Chen Y., and J.-Guo Tian, Sci. Rep. \textbf{3}, 3216 (2013).

\bibitem{Wang2013} L. Wang et al., Science \textbf{342}, 614 (2013).

\bibitem{Gehring2016} P. Gehring,  H. Sadeghi, S. Sangtarash, C-S. Lau, J. Liu, A. Ardavan, J. H. Warner, C. J. Lambert, G. Andrew. D. Briggs, and J. A. Mol,  Nano Lett. \textbf{16}, 4210 (2016).

\bibitem{Varlet2014} A. Varlet, M.-H. Liu, V. Krueckl, D. Bischoff, P. Simonet, K. Watanabe, T. Taniguchi, K. Richter, K. Ensslin, and T. Ihn, Phys. Rev. Letters \textbf{113}, 116601 (2014).

\bibitem{Torres2009} L. E. F. Foa Torres and G. Cuniberti, Apply. Phys. Lett. \textbf{94}, 222103 (2009).

\bibitem{Rocha2010} C. G. Rocha, Luis E. F. Foa Torres, and G. Cuniberti, Phys. Rev. B \textbf{81}, 115435 (2010).

\bibitem{Tien1963} P. K. Tien and J. R. Gordon, Phys. Rev. \textbf{129}, 647 (1963).

\bibitem{Platero2004} G. Platero and R. Aguado, Phys. Rep. \textbf{395}, 1 (2004).

\bibitem{Foa2014} L. E. F. Torres, S. Roche, and J.-C. Charlier, {\it{Introduction to graphene based nanomaterials}}, Cambridge University Press (2014).

\bibitem{Rocha2011} C. G. Rocha, M. Pacheco, L. E. F. Foa Torres, G. Cuniberti, and A. Latg\'e, Eur. Phys. Lett. \textbf{94}, 47002 (2011).

\bibitem{Jhon2010} J. W. Gonz\'alez , H. Santos, M. Pacheco, L. Chico, and L. Brey, Phys. Rev. B {\bf 81}, 195406 (2010).

\bibitem{Jhon2011} J. W. Gonz\'alez, H. Santos, E. Prada, L. Brey, and L. Chico, Phys. Rev. B {\bf 83}, 205402 (2011).

\bibitem{Chico2012} L. Chico, J.W. Gonz\'alez , H. Santos, M. Pacheco, and L. Brey, Acta Physica Polonica A, \textbf{122}, 299 (2012).
 
\bibitem{orellana2013} P. A. Orellana, L. Rosales, L. Chico, and M. Pacheco, J. Appl. Phys. \textbf{113}, 213710 (2013).

\bibitem{liu2009} Z. Liu, K. Suenaga, P. J. F. Harris, and S. Iijima, Phys. Rev. Lett. \textbf{102}, 015501 (2009).

\bibitem{Lee2010} Y. L. Lee, S. Kim, C. Park, J. Ihm, and Y. W. Son, ACS Nano \textbf{4}, 1345 (2010).

\bibitem{haugen2008} H. Haugen, D. Huertas-Hernando, and A. Brataas, Phys. Rev. B \textbf{77}, 115406 (2008).

\bibitem{Wang2015} Z. Wang, C. Tang, R. Sachs, Y. Barlas, and J. Shi, Phys. Rev. Lett. \textbf{114}, 016603 (2015).

\bibitem{DFT} S. Reich, J. Maultzsch, C. Thomsen, and P. Ordej\'on, Phys. Rev. B \textbf{66}, 035412 (2002).

\bibitem{SOB} J. F. Song, Y. Ochiai, and J. P. Bird, Appl. Phys. Lett. \textbf{82}, 4561 (2003).

\bibitem{Parmentier2016} F. D. Parmentier, L. N. Serkovic-Loli, P. Roulleau, and D. C. Glattli, Phys. Rev. Lett. \textbf{116}, 227401 (2016).

\end{thebibliography}
\end{document}